\documentclass[12pt,preprint]{aastex}  
\usepackage{hyperref}

\slugcomment{Draft as of \today.}
\shorttitle{Quasi-Thermal GRB Spectral Models}
\shortauthors{Bellm et al.}

\newcommand{\Ep}{\ensuremath{E_{peak}}}
\newcommand{\Rt}{\ensuremath{{\cal{R}}(t)}}
\newcommand{\R}{\ensuremath{{\cal{R}}}}
\defcitealias{ryde08}{RP08}

\begin{document}
 
\title{RHESSI Tests of Quasi-Thermal Gamma-Ray Burst Spectral Models}

\author{Eric C. Bellm\altaffilmark{1,2}}
\altaffiltext{1}{UC Berkeley Space Sciences Laboratory, 7 Gauss Way,
Berkeley, CA 94720-7450, USA}
\altaffiltext{2}{ebellm@ssl.berkeley.edu}

\begin{abstract}

Prompt gamma-ray burst spectra evolve on short time scales, suggesting that
time-resolved spectral fits may help diagnose the still unknown prompt
emission mechanism.  We use broad-band gamma-ray data from the RHESSI
spacecraft to test quasi-thermal models with high signal-to-noise
time-resolved spectra of nine bright gamma-ray bursts.  In contrast to
results reported in more narrow energy bands, the quality of the fits of
quasi-thermal models is poor in relation to fits of the phenomenological 
Band function.  Moreover, the best-fit parameters for the simplest
quasi-thermal model, a black body plus a nonthermal power law, show
significant dependence on the fit band.  Models that replace the power law
with more complicated nonthermal functions are not robust for the data 
considered here and decrease the physical relevance of the fit black body.

\end{abstract}


%%%%%%%%%%%%%%%%%%%%%%%%%%%%%%%%%%%%%%%%%%%%%%%%%%%%%%%%%%%%%%%%%%%%%%%%%%%%
\section{Introduction} 

One of the most distinct features of the initial gamma-ray emission of
gamma-ray bursts (GRBs) is its temporal variability.  Significant evolution
of the lightcurve and the spectrum occurs on timescales shorter than the
total burst duration.  
Accordingly, spectral fits of subintervals
of a burst may provide improved insight into the burst emission mechanism,
at the cost of increased complexity.

When the \citet{band93} phenomenological spectral model (the ``Band
function'', a smoothly connected broken power law) 
was found to successfully fit the GRB prompt emission observed
by BATSE, systematic time-resolved analyses of large burst samples
focused on identifying patterns in the fit parameter evolution in an
attempt to gain insight into the emission mechanism of long GRBs.  
\citet{ford95}, \citet{crid97}, and \citet{pree98a} considered the evolution 
of the peak energy \Ep, low-energy power-law index $\alpha$, and high-energy 
power-law index $\beta$, respectively.  Broadly, these authors found a typical
hard-to-soft decay of the emission and a general correlation of spectral
hardness with intensity.  

The observation of hard low-energy spectral slopes ($\alpha \sim +1$) in the 
initial portions of GRB pulses raised concerns about the viability of the
synchrotron shock model, as $\alpha > -2/3$ violated the ``line of death''
for optically thin synchrotron (OTS) emission
\citep{crid97,pree98b,ghir02,pree02}.
These violations, in concert with theoretical expectations
from the fireball model \citep[e.g.,][]{mesz00}, 
led authors to suggest such emission
could have a thermal origin \citep{crid97,pree00b}.  \citet{ghir03}
found that time-resolved BATSE LAD spectra with hard low-energy indices 
could be acceptably fit with a black-body spectrum.  
This result was confirmed by
\citet{ryde04} for hard, single pulses.  In some cases the addition of a simple
power-law improved the correspondence with the high-energy LAD data.  These
successes led \citet{ryde05} to propose that all GRB emission might be
decomposed into thermal and nonthermal components of similar magnitude,
with the black-body emission providing the spectral peak of the prompt
emission.
Within the LAD band, a simple power law proved a sufficient approximation
to the more complicated true nonthermal emission, with the resulting
black-body plus power-law model (BBPL) fits having similar $\chi^2$ values
as the Band function for identical degrees of freedom.  In \citet{ryde06},
the authors noted that the power-law slope of the BBPL model avoided the 
OTS line of death and sought to interpret the black body fit results in the
context of a photospheric model.  Most recently, \citet{ryde08}
\citepalias[hereafter,][]{ryde08} 
identified regularity in the time evolution of the temperature and flux of
the black-body component for single-pulse bursts and linked the
normalization of the black-body component to the size of the 
thermal emitting region \citep[see also][]{peer07}.

While the simple power-law approximating the nonthermal emission in the
BBPL model is effective over the moderate energy band (28~keV--1.8~MeV) 
of the BATSE LAD detectors, tests of the BBPL model using data 
covering a broader bandpass have had mixed results.
\citet{mcbr06} successfully fit the BBPL model to GRB 041219A data from
{\it INTEGRAL}-SPI spanning the range 20~keV--8~MeV, although the black body
contributed only a small portion of the flux of the main burst.
\citet{fole08} obtained acceptable time-integrated 
fits of the BBPL model to GRBs observed by {\it INTEGRAL}-IBIS and SPI,
although the IBIS detection requirement created a burst sample with
relatively low peak spectral energies ($\Ep < \sim150$~keV).
\citet{ghir07b} considered BATSE bursts
co-observed in the X-ray by the Beppo-SAX WFC.  Time-resolved spectra were
not available for the WFC data, but a summation of the time-resolved
BBPL fits to the BATSE data led to a significant overprediction of the WFC
flux, while a similar extrapolation of the summed time-resolved Band
function fits was much more successful.

In this work, we have employed data from the Ramaty High Energy Solar 
Spectroscopic Imager (RHESSI) \citep{lin02} to investigate the behavior of
quasithermal spectral models over a broad energy band.  
RHESSI's nine coaxial germanium detectors
image the Sun at X-ray to gamma-ray energies (3~keV--17~MeV) with excellent
resolution in energy (1--5~keV) and time (1 binary $\mu$s) \citep{smit02}.
RHESSI's detectors are effectively unshielded above $\sim$30~keV
and receive emission from astrophysical sources like GRBs with a total
effective area of $\sim$150~cm$^2$.
Each coaxial detector is electronically segmented into thin front and thick
rear segments; most off-axis emission is recorded in the rear segments.
The satellite rotates with a period of about 4 seconds.

%%%%%%%%%%%%%%%%%%%%%%%%%%%%%%%%%%%%%%%%%%%%%%%%%%%%%%%%%%%%%%%%%%%%%%%%%%%%

\section{Data Analysis} \label{sec-analysis} 

\subsection{Sample Selection} \label{sec-sample}

The identification of optimal fit intervals for time-resolved spectral fitting 
requires balancing the need for high signal-to-noise (in order to constrain 
the fit parameters of the spectral models) with the goal of the finest possible
time resolution.  In our previous systematic analysis of RHESSI GRBs
\citep{bellsf08}, we found that a S/N ratio of at least 45 was required
for adequate fitting of the most complicated models.  This value is also
consistent with those adopted by previous studies of time-resolved BATSE
spectra \citep{crid97,kane06}.

Because bursts may have periods of high signal which are degraded to low
S/N when integrated over the whole burst, we chose candidate bursts from
the RHESSI sample\footnote{\url{http://grb.web.psi.ch/}} with localizations
and total S/N $>$ 30.  Radiation damage to RHESSI's germanium detectors has
slowly degraded spectral performance; we restricted our analysis to
detectors which do not show signs of radiation damage. 
We therefore considered bursts from the RHESSI launch in February 2002 through
June 2006, after which all nine RHESSI detectors were damaged.  Forty-five GRBs
met these criteria.

To identify subintervals within each candidate burst, we used 
the Bayesian Blocks algorithm \citep{scar98}, which identifies the most
probable segmentation of a burst lightcurve into intervals of constant
Poisson rate.  We modified the stopping criterion to generate subintervals of
appropriate S/N.  Since the RHESSI data are stored event by event, this
``top-down'' identification of the burst subintervals is more natural than
building up data accumulated on fixed timescales until a minimum S/N is
reached.  While previous work on quasi-thermal models has
focused on spectral evolution within pulses, our segmentation method 
does not require pulse modeling. 
This approach was necessary, as the bright bursts in this
sample generally have irregular temporal structure, but it also avoids
imposing additional implicit selection effects on the burst sample.

We applied the Bayesian Blocks algorithm to the raw RHESSI eventlist to
identify the most probable segmentation points.
The time variation of the background, while present, does not dominate the
GRB variability.  In our modified stopping criterion, we
computed the background-subtracted S/N (in the 60~keV--3~MeV band) for each
proposed subinterval.  If the S/N ratios of both subintervals were
greater than 15, the segmentation was allowed.  For each subinterval, we
applied the Bayesian Blocks algorithm again if the S/N of that interval was
greater than 45.  After segmentation halted, we dropped any leading or trailing
subintervals with S/N $<$ 45 and combined any ``interior'' subintervals with
low S/N with the adjacent subinterval with lower S/N.  The resulting
subintervals all have S/N $>$ 45.  We conducted spectral fitting only on
bursts with at least three subintervals.

\begin{figure} 
\plotone{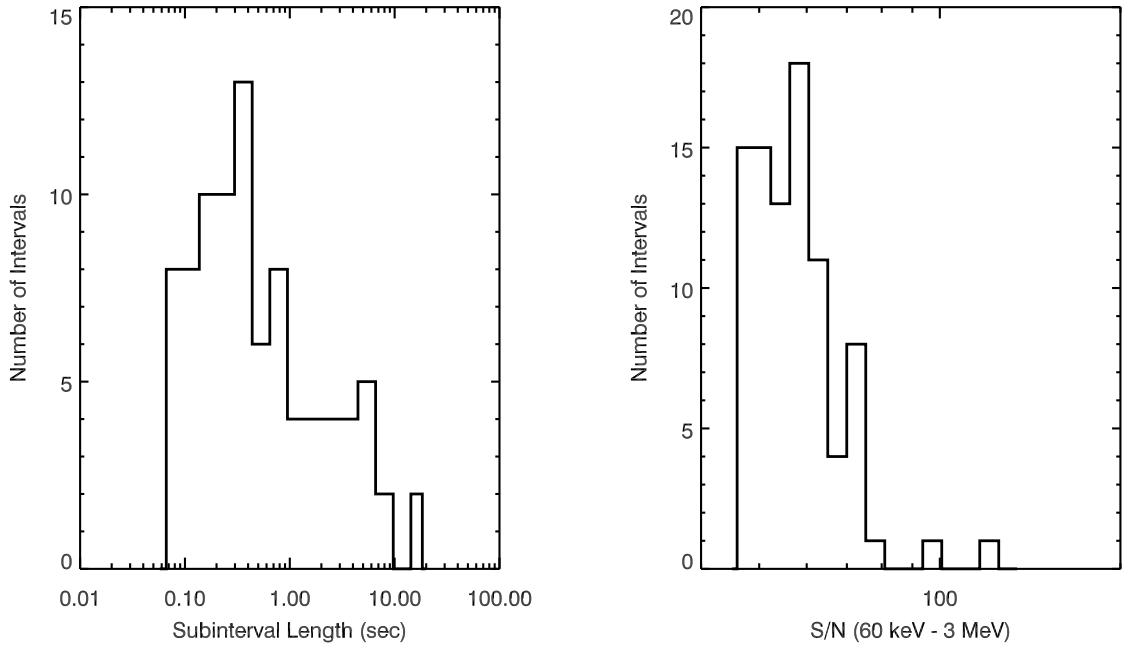}
\caption{
Distribution of subinterval length and signal-to-noise ratio for the 
high S/N burst sample considered in this work.
}
\label{fig-tobs_sn_hist} 
\end{figure}

The resulting burst sample consists of nine bursts and a total of 88
subintervals.  Table \ref{tab-intervals} 
lists the bursts and intervals
considered in this work.  Figure \ref{fig-tobs_sn_hist} shows the distribution 
of subinterval length and S/N.  These bursts are predominantly from early in
the RHESSI mission because for those bursts a greater number of the
detectors remained unaffected by radiation damage.

In order to investigate the spectral evolution of single GRB pulses in
Section \ref{sec-evolution}, it was
necessary to set a lower S/N threshold.  We used our modified Bayesian
Blocks routine to identify subintervals with S/N $>$ 25 within single,
separated pulses.  This ``single-pulse'' sample yielded 25 subintervals
within 3 GRBs (Table \ref{tab-intervals}).

\subsection{Spectral Fitting}

\subsubsection{Fit Methods}

The RHESSI data were extracted in
SSW-IDL\footnote{\url{http://www.lmsal.com/solarsoft/}}.  
We fit a third-order polynomial
to background intervals immediately before and after the burst, allowing
for a potential periodic modulation of the background with the RHESSI
rotation.  We selected energy binning for the spectra by requiring at least
ten counts in each bin of the raw source spectrum.  We used the same energy
bins for all subintervals of each burst.  Both the source and
background spectra covered the energy range 30~keV--17~MeV.

We determined RHESSI's spectral response by simulating monoenergetic
photons impinging on a detailed mass model in the Monte Carlo suite MGEANT
\citep{stur00}.  For a grid of angles relative to the spacecraft rotation
axis, we simulated photons along 60$^\circ$ arcs in rotation
angle.  After interpolating to the appropriate off-axis angle, we weighted
the 60$^\circ$ sector responses by the burst light curve to generate the
total response.  Since many of the subintervals considered in this work
were shorter than one-sixth of RHESSI's four second rotation period,
responses for short intervals were often single sector responses.  Our
simulations indicate that the spectral response does not vary appreciably
on shorter scales in rotation angle.

We utilized ISIS v1.4.9 \citep{citeisis} for spectral fitting.
Fitting was automated via an ISIS script and the results stored in a database.

We used data from the RHESSI rear segments only for all bursts except for
GRBs 021206 and 021008A, which had off-axis angles small enough
(18$^\circ$ and 50$^\circ$, respectively) 
that appreciable counts were recorded in the
front segments.  (Coincidentally, these bursts also both had data decimation
in the rear segments---only a fraction of the observed counts were
recorded.)
The fit energy band was 30~keV--17~MeV, with some
exceptions:  GRB 021206 had significant atmospheric backscatter in the rear
segments \citep{wigg08}, 
so the front and rear segment data were kept separate and fit
over the range 30~keV--2800~keV and 300~keV--17~MeV.  (For all other
bursts, data from all detectors were combined into a single spectrum.)
GRBs 030519B and 040228 came through the extreme rear of the RHESSI
cryostat (166$^\circ$ and 163$^\circ$), where the low-energy response
is more sensitive to the detailed mass modelling, so a fit range of
50~keV--17~MeV was employed.  Finally, GRB 021008A required a fit band of
100~keV--17~MeV for adequate fitting; \citet{wigg08} found similar behavior
in a time-integrated fit of this data.

\subsubsection{Spectral Models}

We fit both the time-resolved and the time-integrated spectra with a
variety of spectral models commonly used to represent the GRB prompt
emission.  The Band function (Band) is a phenomenological model which fits the
vast majority of broadband prompt spectra \citep{band93}. 
It is a broken power law with a smooth transition between the upper and
lower power laws:
\begin{equation}
N_E = \left\{
\begin{array}{lr}
  A (E/E_{piv})^{\alpha} \exp(-E (2+\alpha)/\Ep) & E<E_{break} \\
  B	(E/E_{piv})^{\beta}               & E>E_{break}
\end{array}
\right.
\end{equation}
with $E_{break} \equiv E_{0} (\alpha - \beta)$ and 
$B \equiv  A (\frac{(\alpha-\beta) E_0}{E_{piv}})^{\alpha-\beta}
\exp(\beta-\alpha)$.  For $\beta < -2$ and
$\alpha > -2$, \Ep\ corresponds to the peak of the
$\nu F_\nu$ spectrum.  The normalization $A$ has units of photons cm$^{-2}$
s$^{-1}$ keV$^{-1}$, and $E_{piv}$ is here taken to be 100 keV. 

While the Band function is often assumed to be the ``true'' spectral model
in fits of GRB spectra, data obtained over a more limited energy band or at
lower S/N may not be sufficient to constrain its parameters.  We 
also tested other empirical models with fewer free parameters.  
The cutoff power law (CPL) is equivalent to the Band function below $E_{break}$:

\begin{equation}
N_E = A (E/E_{piv})^{\alpha} \exp(-E (2+\alpha)/\Ep)
\end{equation}

We also fit a single power law (PL): $N_E = A (E/E_{piv})^{\alpha}$.  While
the PL model was a poor fit to these spectra, its fit spectral index was
useful in comparison with the other models.

%An $E_{piv}$ of 1~keV was used for both the CPL and the PL models.  While
%previous work has found that this low pivot value introduced unneccessary
%covariance in the fit parameters for GRB data \citep{}, we do not fit the
%normalization $A$ directly.  Rather, we use ISIS
%to tie the model normalization to a function
%which computes the model fluence over a chosen energy band.  This fluence
%is then varied like the other parameters during the minimization to find
%the best fit.  We also compute the confidence limits on the fluence
%directly by stepping through a fluence grid and refitting the other fit
%parameters while monitoring the change in $\chi^2$.  The fluence confidence
%limits so obtained avoid the difficult error propagation of the other
%covariant fit parameters.  

The simplest quasi-thermal spectral model is a black body plus power law
model (BBPL).   Proposed by \citet{ryde05},
it consists of Planck function and a simple power law:

\begin{equation}
N_E = A \frac{(E/E_{piv})^2}{\exp{E/kT} - 1} + B (E/E_{piv})^s
\end{equation}

The approximation of the nonthermal emission by a simple power-law is an
approximation which is generally sufficient over the BATSE energy band
\citep{ryde05}.  However, it is not expected to fit over a broader band
\citep{ryde04,ryde05,ryde08}.  Like the PL model, it diverges unphysically.

More sophisticated quasi-thermal models replace the nonthermal PL in the 
BBPL with an empirical function with more free parameters.  
We attempt to fit a black body plus cutoff power law (BBCPL) model as well:

\begin{equation}
N_E = A \frac{(E/E_{piv})^2}{\exp{E/kT} - 1} + B (E/E_{piv})^{\alpha} 
	\exp(-E (2+\alpha)/\Ep)
\end{equation}

(We retain the notation of \Ep\ for the cutoff parameter of the BBCPL model
for clarity of comparison, but in some cases the true peak of the $\nu F_\nu$
spectrum may be due to the black-body component at $3.92~kT$.)
However, the addition of more free fit parameters 
leads to convergence problems even with our relatively high S/N 
broad-band spectra.  In addition to the unconstrained fits of the BBCPL 
model, then, we also perform fits in which the variation of
the model temperature is constrained to an expected range.
We describe the specifics of this constrained
quasi-thermal model in the text below.

Given the convergence problems faced by the BBCPL model, we have not
attempted fits of a black body + Band model, which contains an additional
free parameter.  
In most cases, the simple CPL model provides an adequate representation of the
RHESSI data, with the Band model offering only moderate improvements.
Accordingly, the BBCPL model should provide a sufficient test of the
importance of the black-body component over the RHESSI band.

%%%%%%%%%%%%%%%%%%%%%%%%%%%%%%%%%%%%%%%%%%%%%%%%%%%%%%%%%%%%%%%%%%%%%%%%%%%%
\section{Fit Results} \label{sec-fits}

\begin{figure} 
\plotone{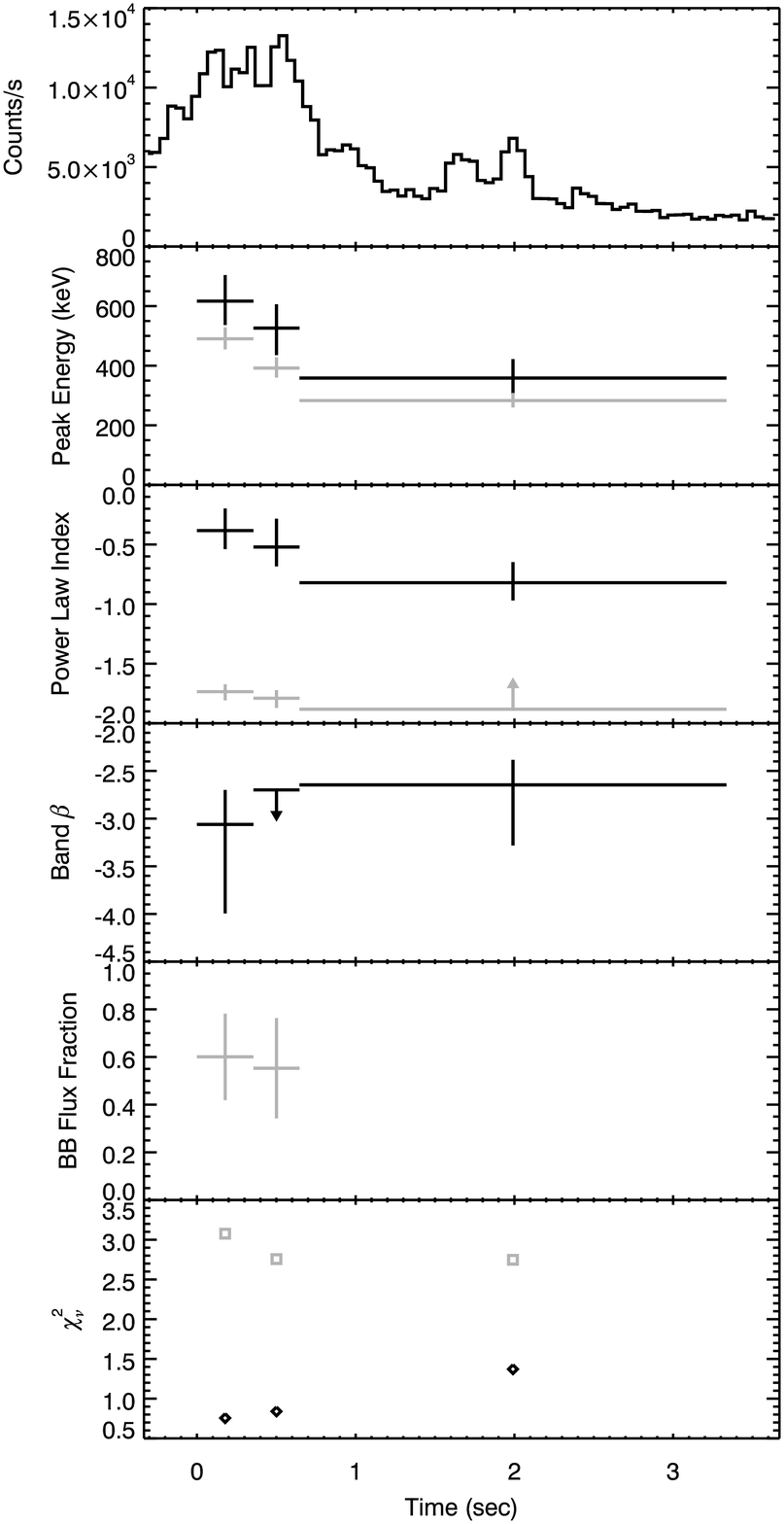}
\caption{
Spectral evolution of GRB 020715.  The 60~keV--3~MeV 
lightcurve has 0.05~s time bins; counts are shown only for those segments
used in the fits.  The fit model parameters are black for the Band model
and gray for the BBPL model.  All errors are 90\% C.L.  
Peak energies are \Ep\ for the Band model and $3.92~kT$ for the BBPL model.
The Band model $\alpha$ is plotted with the power-law index
$s$ of the BBPL model.
The errors on the fraction of the flux provided by the black body in the 
RHESSI band were estimated from a Monte Carlo bootstrap.
Points where the value was unconstrained in the fit are omitted from the
plot.
}
\label{fig-020715} 
\end{figure}

\begin{figure} 
\plotone{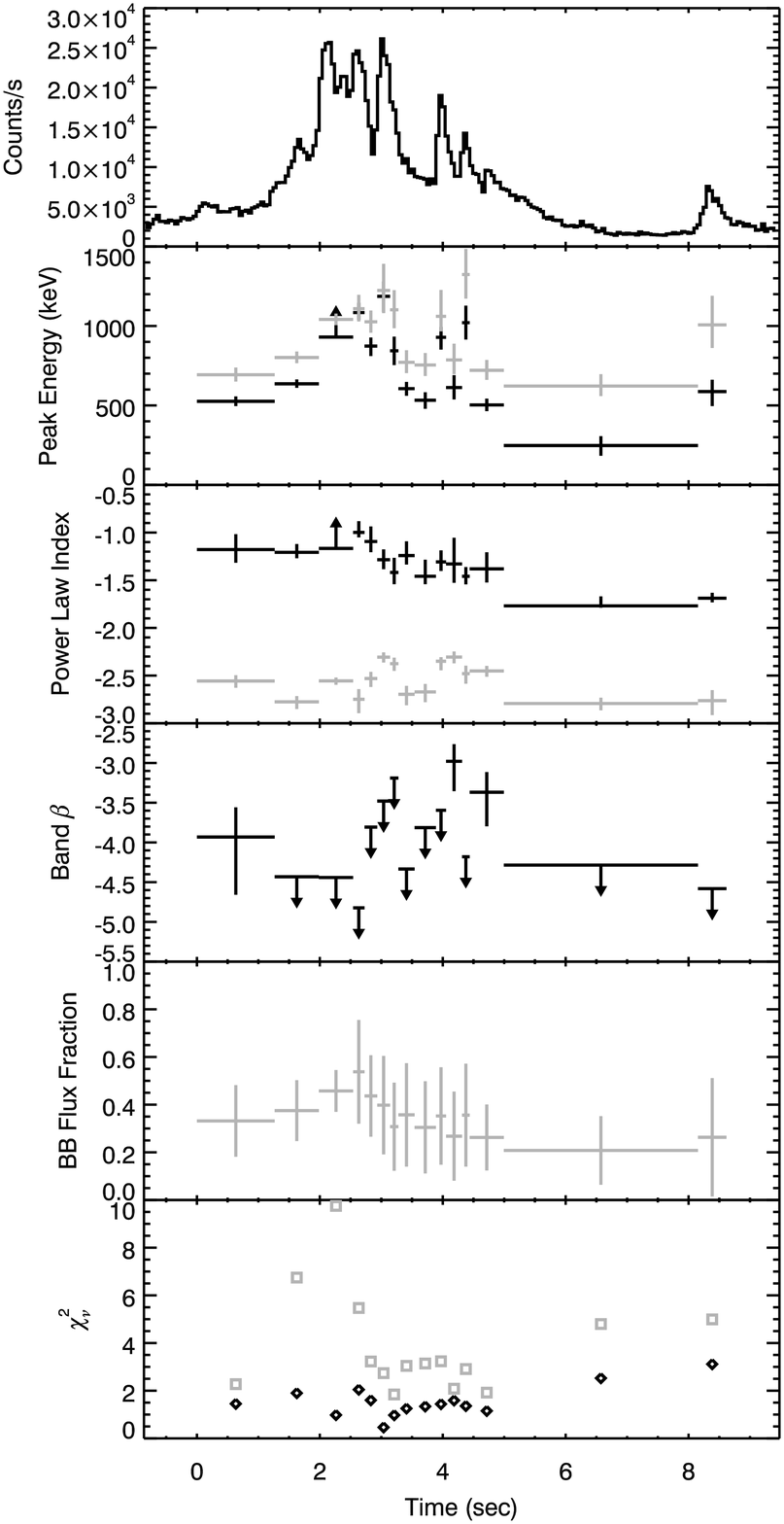}
\caption{
Spectral evolution of GRB 021008A.  Symbols as in Fig. \ref{fig-020715}.  
}
\label{fig-021008A} 
\end{figure}

\begin{figure} 
\plotone{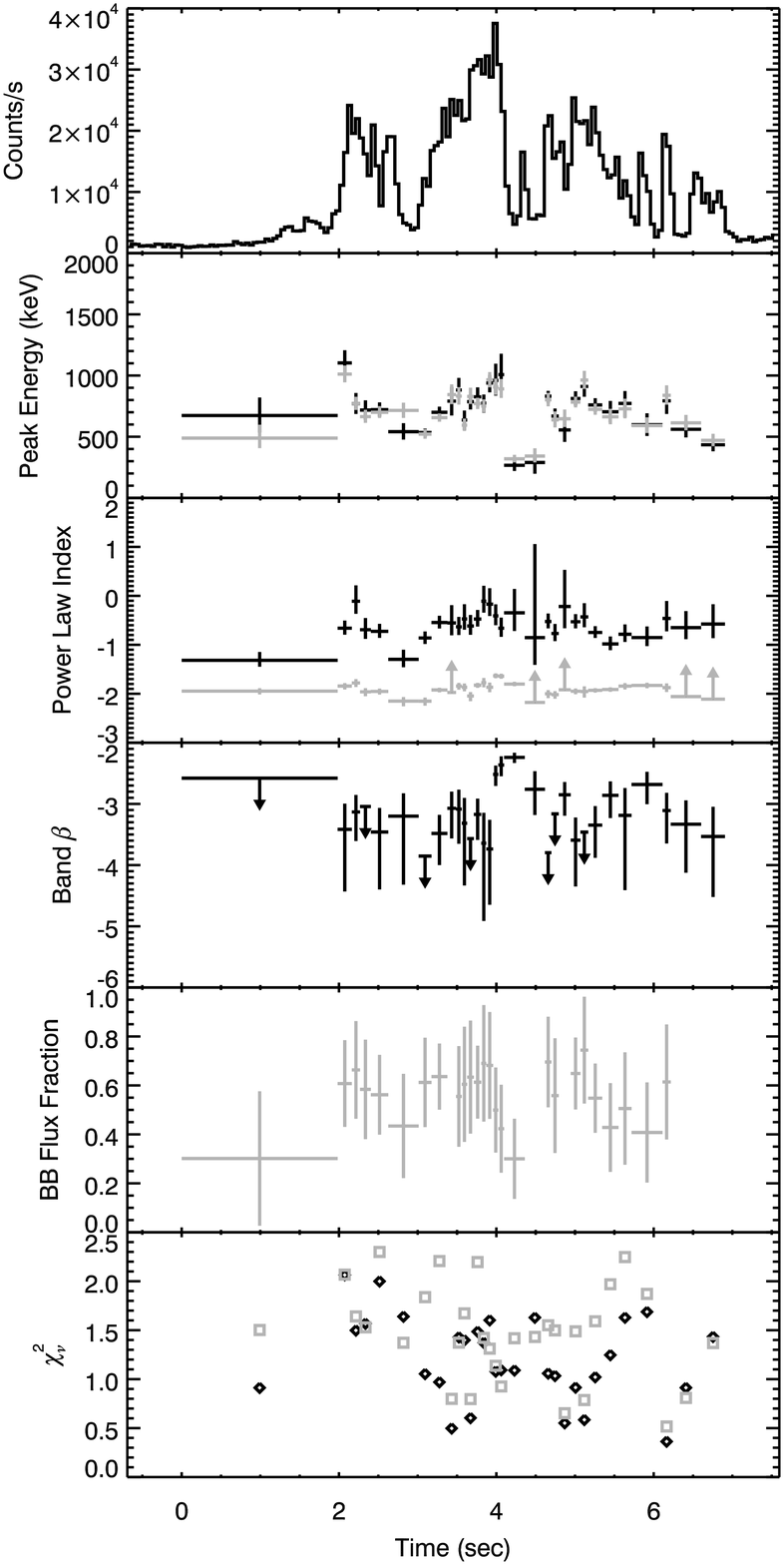}
\caption{
Spectral evolution of GRB 021206.  Symbols as in Fig. \ref{fig-020715}.  
}
\label{fig-021206} 
\end{figure}

\begin{figure} 
\plotone{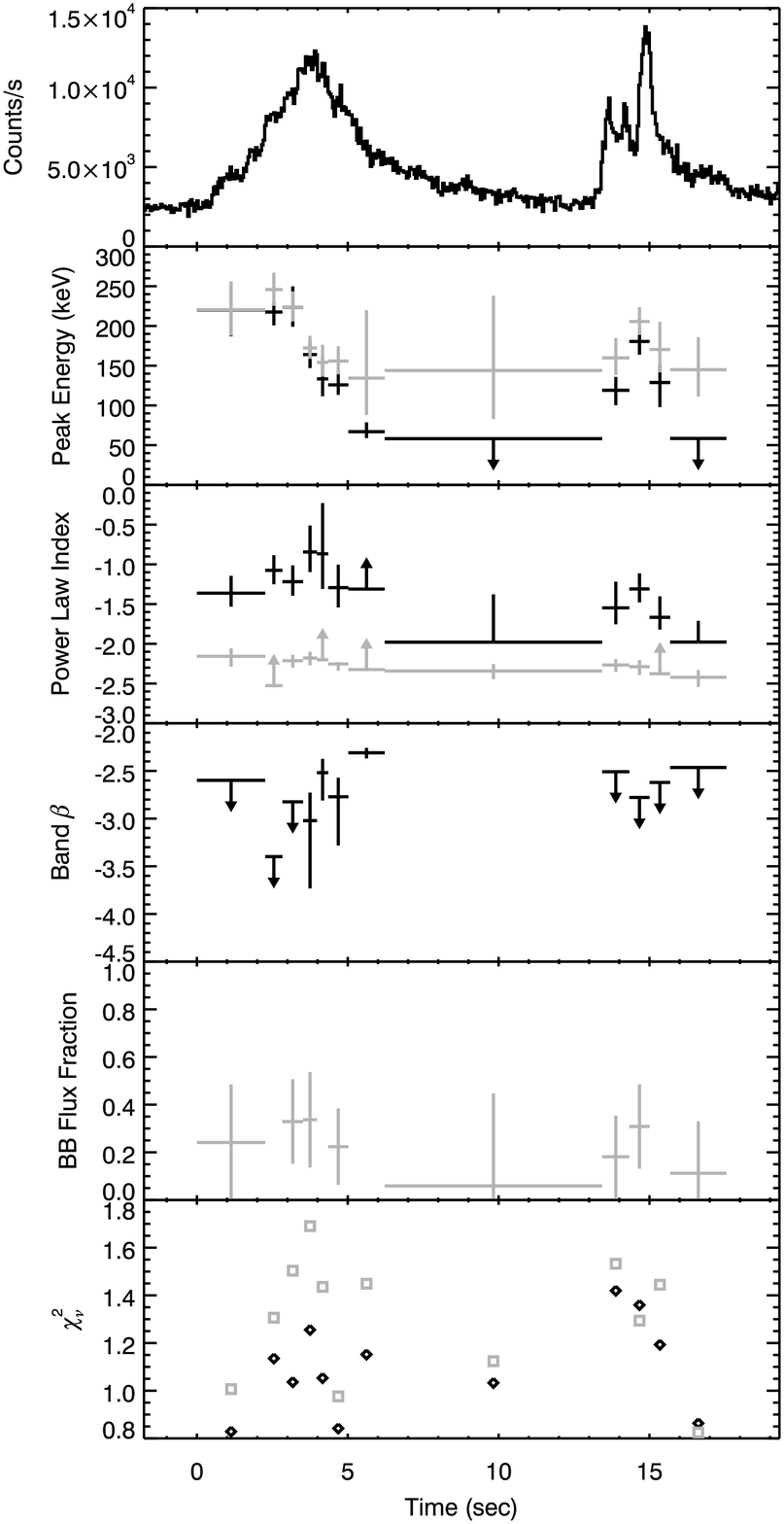}
\caption{
Spectral evolution of GRB 030329.  Symbols as in Fig. \ref{fig-020715}.  
}
\label{fig-030329} 
\end{figure}

\begin{figure} 
\plotone{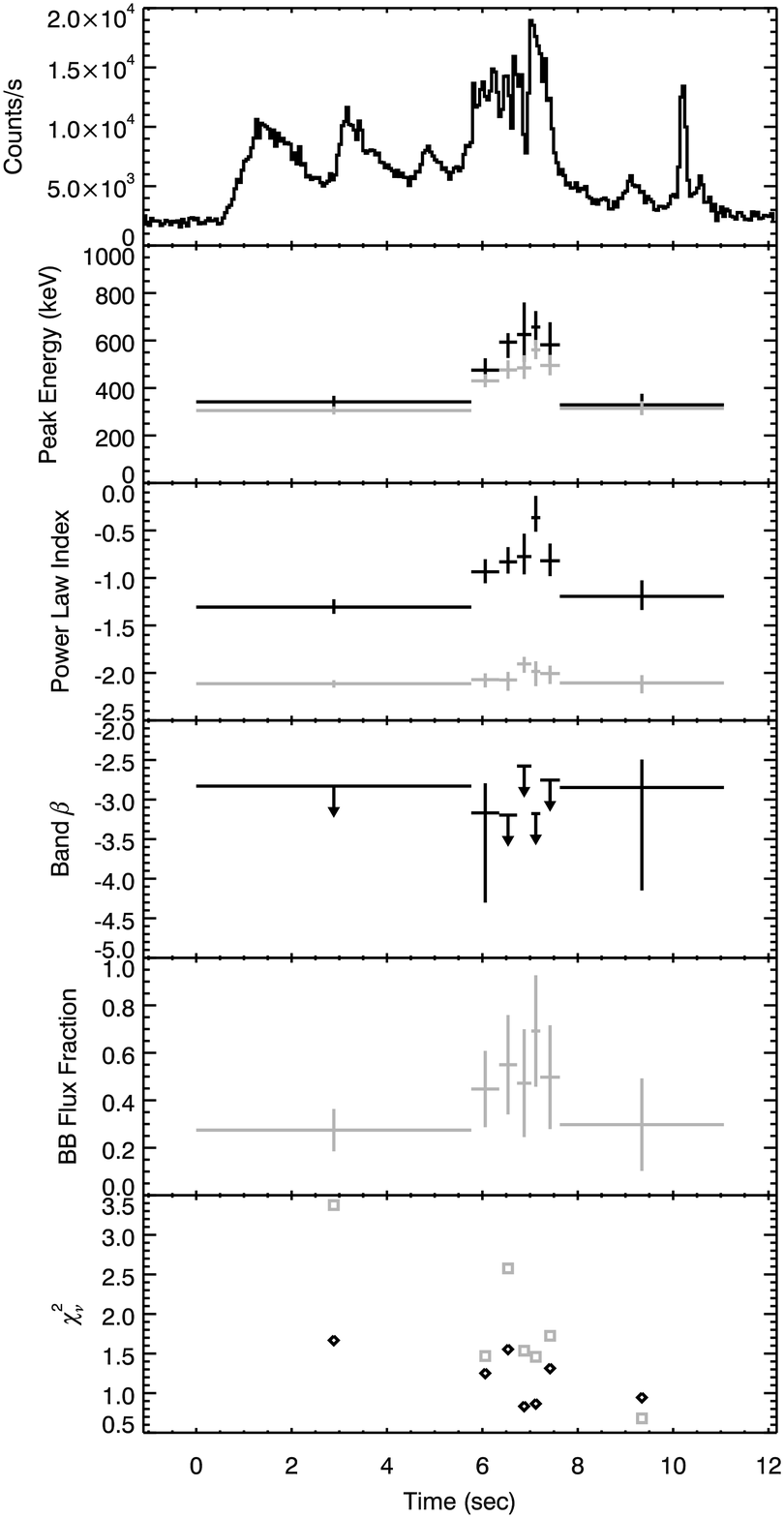}
\caption{
Spectral evolution of GRB 030519.  Symbols as in Fig. \ref{fig-020715}.  
}
\label{fig-030519} 
\end{figure}

\begin{figure} 
\plotone{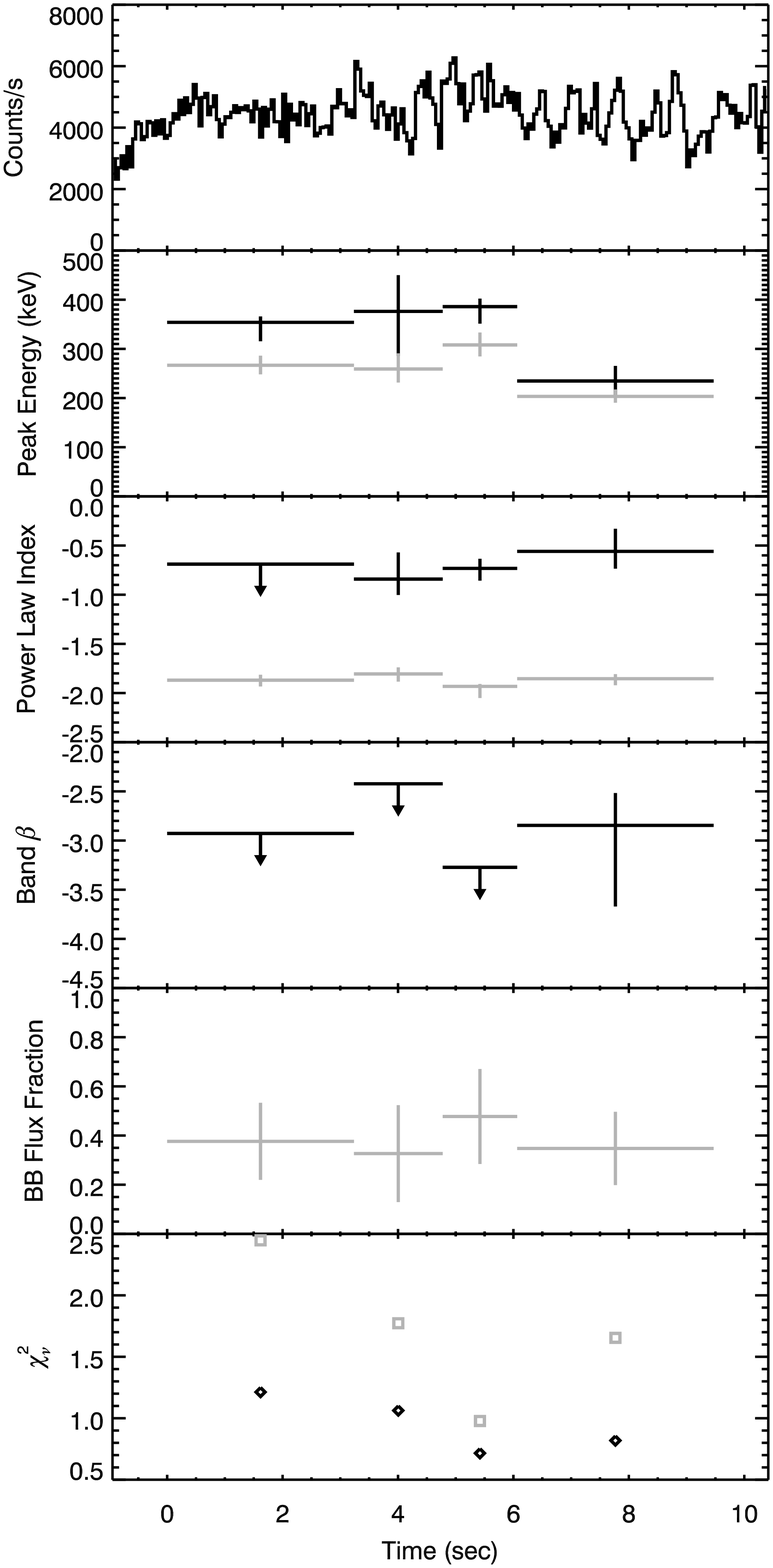}
\caption{
Spectral evolution of GRB 031027.  Symbols as in Fig. \ref{fig-020715}.  The
lightcurve appears flat because the background level is not visible in
the time interval plotted.
}
\label{fig-031027} 
\end{figure}

\begin{figure} 
\plotone{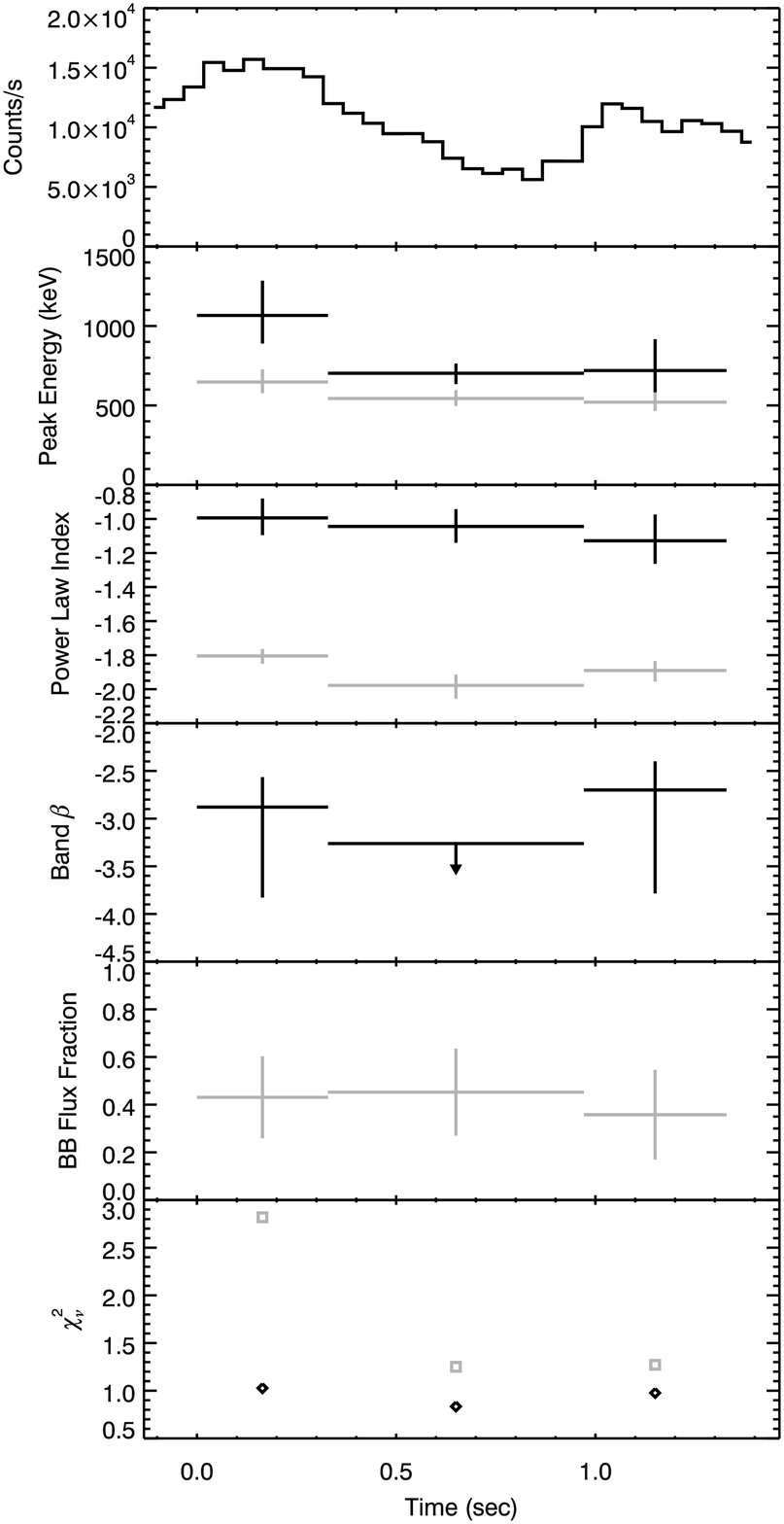}
\caption{
Spectral evolution of GRB 031111.  Symbols as in Fig. \ref{fig-020715}.  
}
\label{fig-031111} 
\end{figure}

\begin{figure} 
\plotone{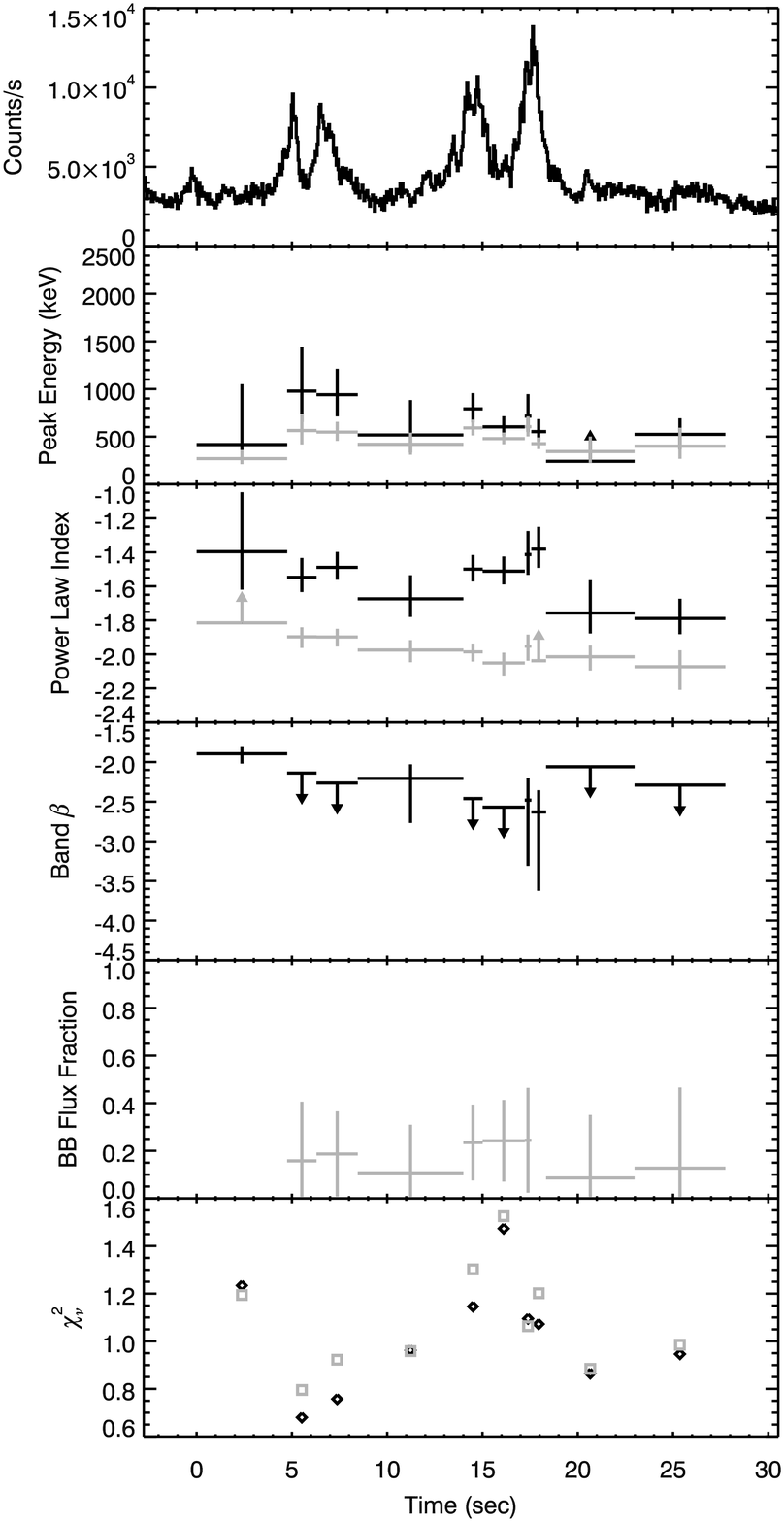}
\caption{
Spectral evolution of GRB 040228.  Symbols as in Fig. \ref{fig-020715}.  
}
\label{fig-040228} 
\end{figure}

\begin{figure} 
\plotone{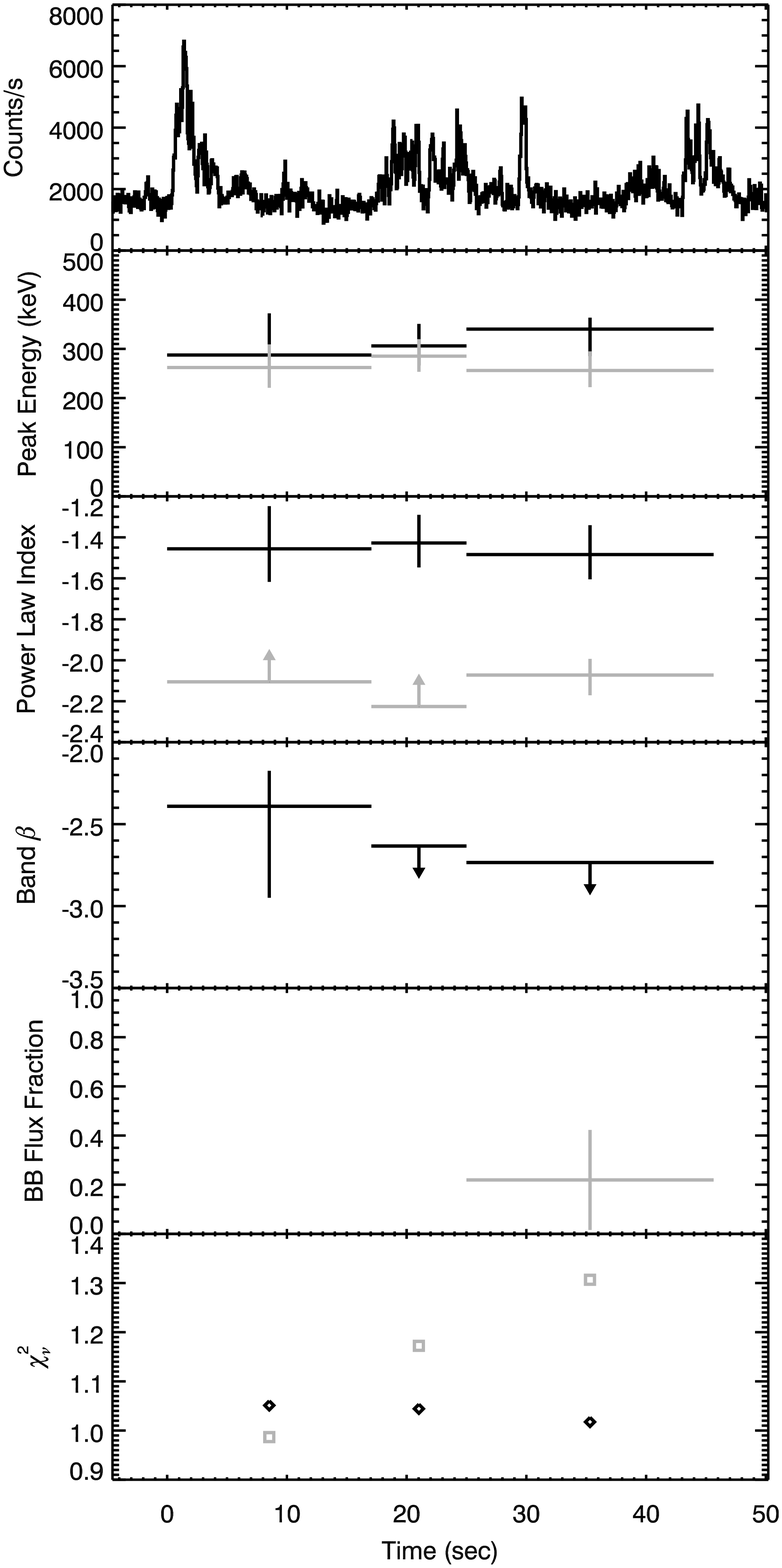}
\caption{
Spectral evolution of GRB 040810.  Symbols as in Fig. \ref{fig-020715}.  
}
\label{fig-040810} 
\end{figure}

We present the time-evolution of the fit parameters of the Band and BBPL
models in Figures \ref{fig-020715}--\ref{fig-040810}.
The quality of the fits 
and the constraint of their parameters were generally good.  For GRB
021206, the goodness of fit was poorer for some subintervals because of the
separation of front and rear data.  Slight offsets in the background
subtraction created disagreements about the overall model normalization and
increased the $\chi^2$.  We elected to take the averaged fluence value
produced by the fit rather than introduce a fit normalization offset
between the front and rear data for this burst.  For GRB 021008A, the only
other burst with front segment data, our combination of front and rear
segment spectra averaged out such background subtraction problems
and obviated the need for such offsets.

\citet{wigg08} fit RHESSI data for many of
these bright bursts using an identical mass model but independent analysis
and fit procedures.
Our fits of the total burst intervals are a good match to those results 
(Table \ref{tab-fit_comps}).  In the case of GRB 021206, the presence of
excess high-energy emission not well-modeled by the Band function increased
the $\chi^2$ value for the fit to 17~MeV.  (This component likely also
contributed to the poor goodness of fit for this burst 
in the time-resolved fits, as
discussed above, although its impact is harder to ascertain.)
{\it HETE}-2 also observed three bursts in this sample.  RHESSI fit higher
\Ep\ values for all three bursts; in two of the cases, the peak energies
were above the 400~keV upper bound of the {\it HETE} energy band.

\begin{figure} 
\plotone{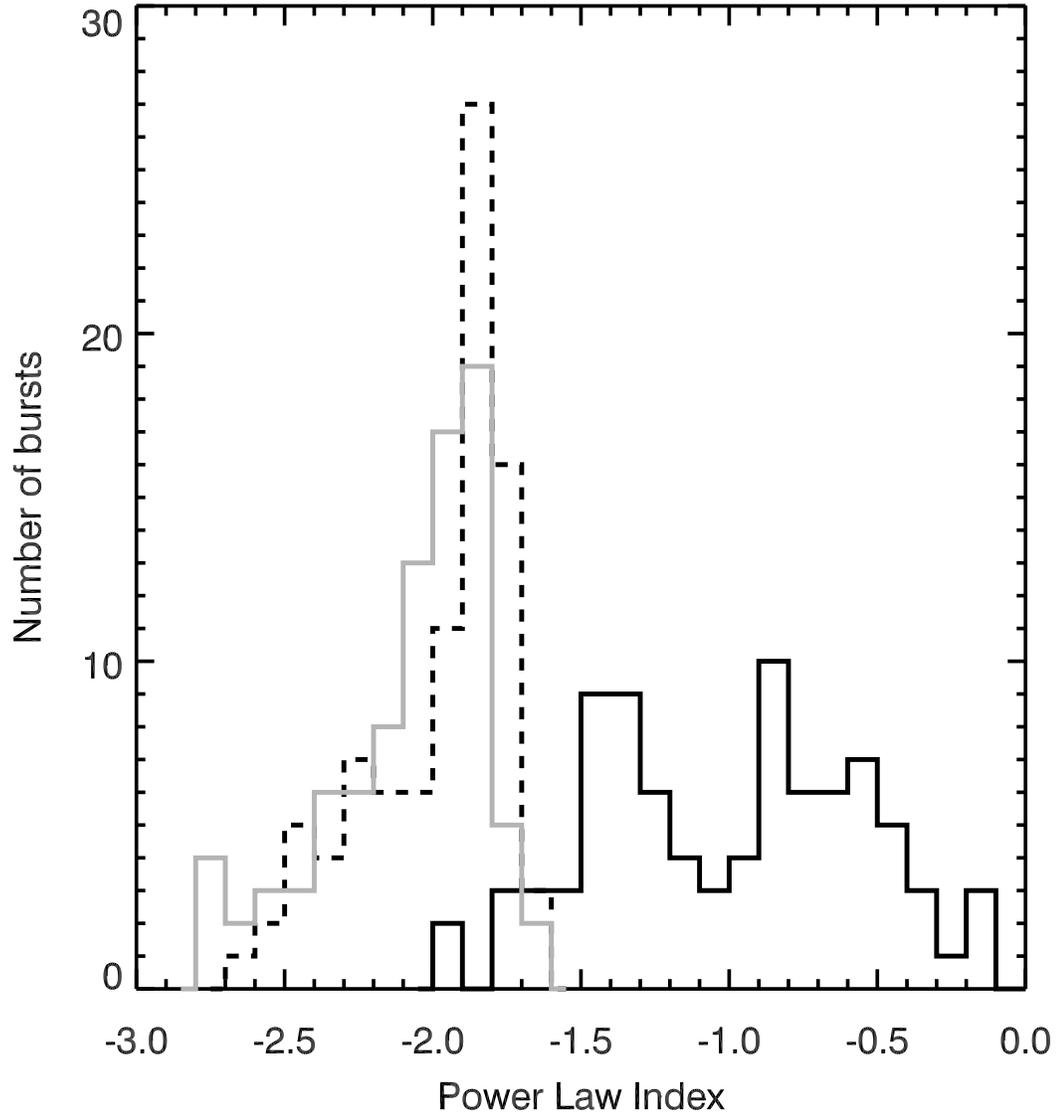}
\caption{
Histogram of the fit low-energy power law indices for the Band model (solid
black), BBPL model (solid gray), and simple power-law model (dashed black).
The indices of the PL model and the BBPL model are highly correlated.
The apparent bimodality of the Band $\alpha$ index is likely an artifact of
the limited burst sample (Section \ref{sec-fits}).
}
\label{fig-alpha_hist} 
\end{figure}

The distributions of our best-fit parameters are in general agreement
with the BATSE results obtained by \citet{kane06}.  For our time-resolved
Band function fits, the mean $\alpha$ is $-1.00$ with standard deviation
$0.47$ (Figure \ref{fig-alpha_hist}).  
The RHESSI mean \Ep\ is 580 $\pm$ 280~keV, and $\beta$ has mean
$-3.50$ and standard deviation $0.89$.  The BATSE results for all good Band
%fits are $\alpha = -0.90^{+0.23}_{-0.26}$, $\Ep = 268^{+127}_{-98}$~keV,
%and $\beta = -2.33^{+0.26}_{-0.32}$, where the ranges are quartile
%dispersions.
fits are $\alpha = -0.90^{+0.34}_{-0.39}$, $\Ep = 268^{+188}_{-145}$~keV,
and $\beta = -2.33^{+0.39}_{-0.47}$, where we have converted the quartile
dispersions to $1\sigma$ uncertainties under the simplifying assumption of
an underlying normal distribution.
Using Levene's test, we find that the variences of the BATSE and RHESSI
distributions are consistent.  Since the RHESSI distributions are not
normal, we cannot compare the means using Student's $t$-test.  The
nonparametric Kolmogorov-Smirnov test rejects the null hypothesis
BATSE and RHESSI parameter samples are drawn from the same distribution.
However, this result is not surprising, 
given the different fit energy ranges and
the few RHESSI bursts fit here.  Of the three parameters, the alpha
distributions are most consistent despite the apparent bimodality of the
RHESSI sample (Fig. \ref{fig-alpha_hist}).

%%%%%%%%%%%%%%%%%%%%%%%%%%%%%%%%%%%%%%%%%%%%%%%%%%%%%%%%%%%%%%%%%%%%%%%%%%%%
\section{Evaluating Quasithermal Models} \label{sec-therm}

We did not attempt to determine the best-fit model for each spectrum fit in
this work.  With the exception of the Band function and the BBPL, the
models fit herein have different numbers of fit parameters.
As discussed by \citet{prot02},
the conventionally used $F$-test of an additional term
is inappropriate if the models being compared are not nested (i.e.\ the fit
parameters of the parsimonious model are not a subset of the more
complex model).
The rigorous alternative---Bayesian hypothesis testing---would require
numerical marginalization over our fit parameters and hence a major reworking 
of our analysis procedures.  Instead, we have opted to evaluate the
effectiveness of the quasi-thermal models in three ways: first, by directly
comparing goodness-of-fit for those pairs of quasi-thermal and empirical
models with identical degrees of freedom; second, by assessing the
physical plausibility of the model fits; and third, by considering the
robustness of the fits and their relative consistency with each other and
previous work.  

\subsection{BBPL}

We begin by considering the simplest quasi-thermal model, the black body +
simple power law (BBPL).  
This model is not expected to be effective over a broad energy band
\cite[e.g.,][]{ryde05,ghir07b}.  
It is instructive to ask, however, how the properties of the BBPL
fits change when fit over a broader band.  In particular, evidence of
band-dependence calls into question the universality of results
obtained even in a narrow band centered on the peak of the prompt emission.

\begin{figure} 
\plotone{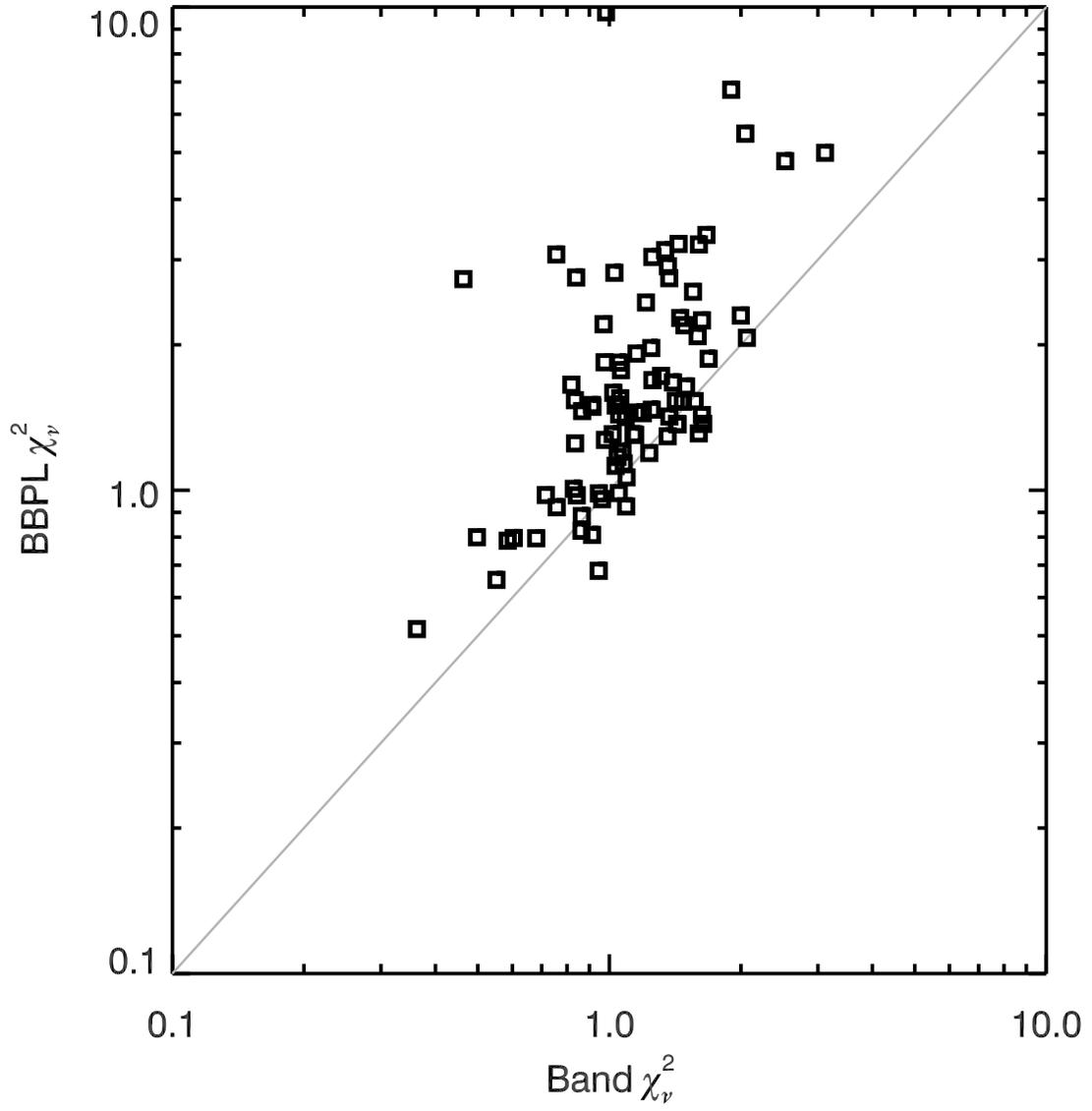}
\caption{
Comparison of the reduced chi-squared values for fits to the Band and BBPL
models, which have identical degrees of freedom.  The Band model is
generally preferred.
}
\label{fig-chisq} 
\end{figure}

The Band and BBPL models each have four fit parameters ($\alpha, \Ep,
\beta$, and the normalization for Band; $kT$, $s$, and the two
normalizations for the BBPL).  Comparing their $\chi^2$ values for each
subinterval (see Fig. \ref{fig-chisq}), 
we find that in 73 of 88 cases, the Band function has a lower
$\chi^2$ value and is thus statistically favored.  

\begin{figure} 
\plotone{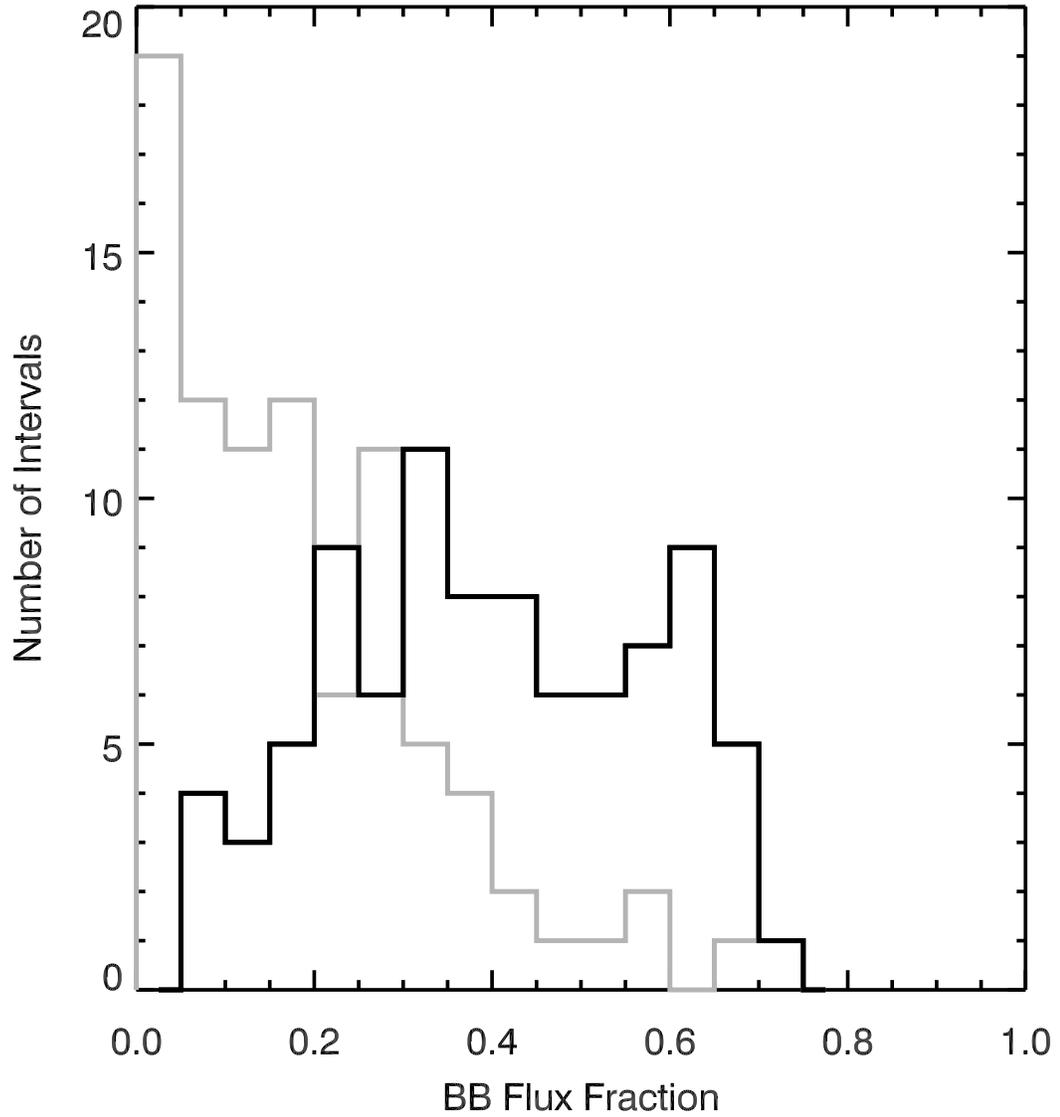}
\caption{
Histogram of the proportion of flux in the RHESSI band contributed by the
black-body component in the BBPL (black) and BBCPL (gray) models.  With the
nonthermal component providing the spectral break in the BBCPL model, the
relevance of the black body component diminishes.
}
\label{fig-bbfrac_hist} 
\end{figure}

As in previous studies, we found that the peak of the blackbody emission
($3.92~kT$) in BBPL fits corresponded closely to the value of \Ep\ obtained
in Band fits (Figure \ref{fig-epeaks}b).  Moreover, the contribution
of the BBPL black-body flux to the overall flux in the band was significant
(Figure \ref{fig-bbfrac_hist}).

Contrary to our expectation, we did not find systematic deviation in the
residuals of the fit BBPL model at low or high energy.  Such deviation
would be symptomatic of the need for additional spectral components 
\citep[e.g.,][]{ryde08}.  

RHESSI's broader energy band did affect the best-fit value of $s$, the
power law index of the BBPL model.  The distribution of $s$ reported by
\citet{ryde06} peaks around -1.5, while our histogram of $s$ peaks near -1.9
(Fig. \ref{fig-alpha_hist}).  This shift may be due to the small sample
of bursts fit here, 
although as discussed in Section \ref{sec-fits} the fit Band index $\alpha$
shows only a small shift compared to the BATSE sample.

However, a softer fit index is expected when fitting a
power law to GRB data extending to higher energies, as GRB spectra
generically fall off more sharply above \Ep.  The fit single power law index 
is thus a band-dependent average of the more complicated spectral shape.  
Indeed, the distribution of $s$ obtained is quite similar to that of the
index of a simple power-law fit to these data (Figure
\ref{fig-alpha_hist}), and $s$ is highly correlated with the fit PL index
for individual intervals.  %possible figure
Since the asymptotic spectral index must be greater than -2 below the peak
energy and less than -2 above it, fitting a power law to a band of
sufficient width which includes the peak energy will average to an
intermediate value.

One of the strengths of the BBPL model is that its nonthermal component
typically is softer than the ``line of death'' of $-2/3$ predicted for
optically thin synchrotron emission.
However, this argument appears to be weakened by the sensitivity of the 
fit power-law index to the range of the data above the peak energy.
It is not clear that the fit value of $s$ provides useful insight into the
nonthermal emission physics, as its value may be an artifact of the 
fit band.

\subsection{BBCPL and Variants}

Given the challenges faced by the BBPL model in the RHESSI band, we
tried more complicated nonthermal components in an attempt to find a
functional broad-band quasithermal model.  We sought to evaluate the
efficacy and robustness of such models and to assess whether the thermal
component retained physical relevance and realism.

After a simple power-law, the cutoff-power law (CPL) model is the next most
basic nonthermal model.  Unlike the PL model, it has the advantage of 
providing a generally adequate representation of the RHESSI data
considered here over the full band.  Accordingly, we tested a black-body plus
cutoff power law (BBCPL) model.

\begin{figure} 
\plotone{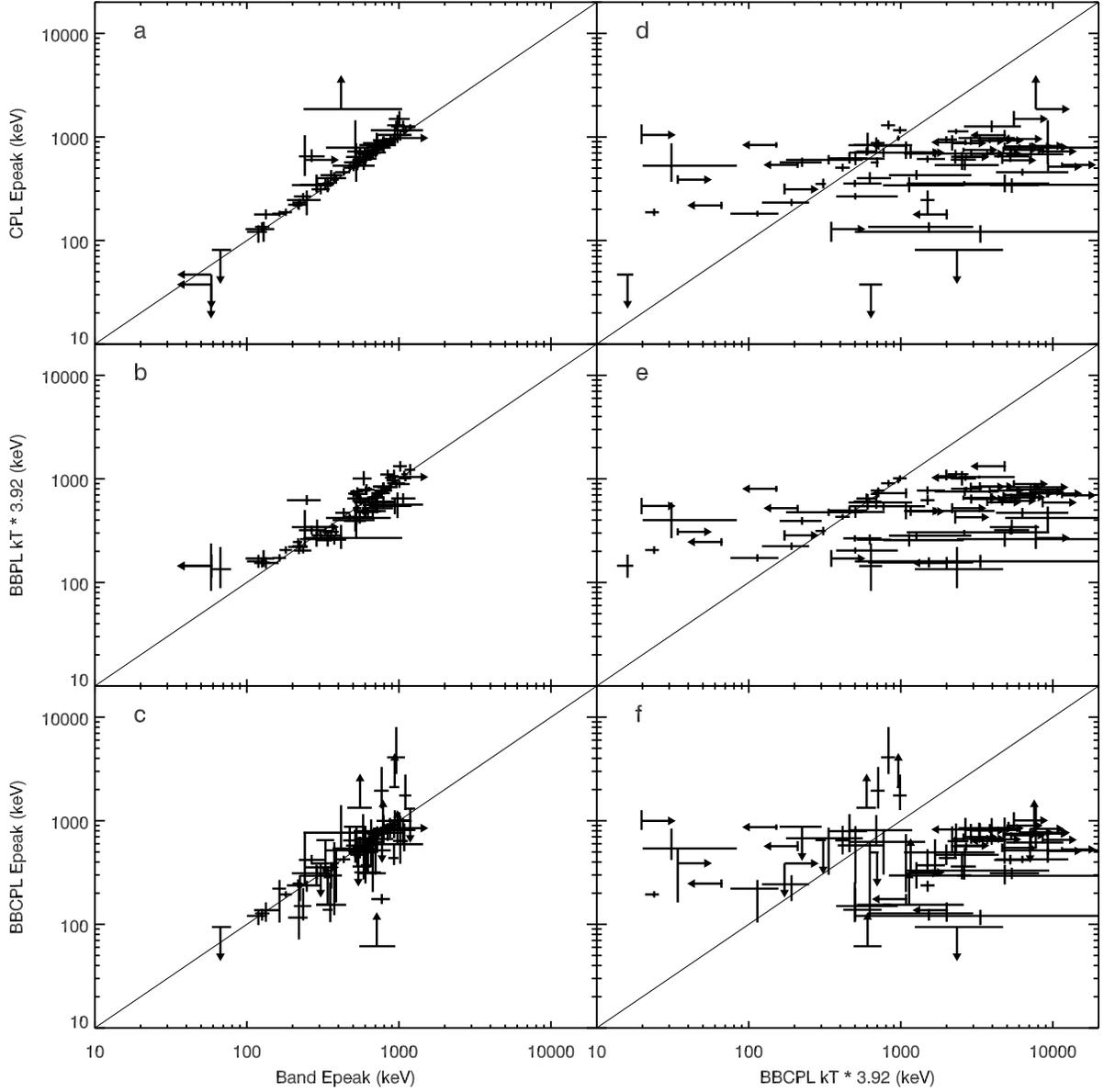}
\caption{Comparison of peak energies obtained in Band, CPL, BBPL, and BBCPL
fits.  As expected, there is good agreement between the \Ep\ values of the
Band and CPL models (panel a).  
Notably, the black-body peak ($3.92~kT$) of the BBPL
model and the BBCPL \Ep\ parameter also match the Band function 
\Ep\ (panels b and c).
However, the black-body peak of the BBCPL model shows little correlation
with the spectral peak identified by the other models (panels d--f).
A few points which were completely unconstrained in one dimension are
omitted in the plot.
}
\label{fig-epeaks} %the label has to go here or it doesn't work!
\end{figure}

Given the success of the basic CPL model at fitting these data, it is
perhaps not surprising that the CPL component of the BBCPL model dominates
the fits.  Figure \ref{fig-epeaks}c shows that the \Ep\ of the BBCPL
fits closely corresponds to the peak energy fit by the Band function.  The
peak energy of the black-body component alone, 
however, is almost completely uncorrelated
with the peak energy of the phenomenological models (Figures
\ref{fig-epeaks}d--f).  This behavior is in
marked contrast to the success of the BBPL model of identifying the
spectral peak with the black-body peak.  Moreover, many of the fit values
of $kT$ in the BBCPL model are limits only, indicating the lack of
robustness in the fits.

The bolometric black-body flux is obtained by integrating the differential
energy spectrum of the Planck function $E N_E$ over all energies; 
it equals $A (kT)^4 E_{piv}^2 \pi^4/15$.
Its value is fairly similar for the BBPL and
BBCPL models. However, since the best-fit black-body temperature is often near 
the limits of the RHESSI band, the fraction of the flux in the
RHESSI band provided by the black body drops sharply in the BBCPL model
(Figure \ref{fig-bbfrac_hist}).
The black-body component is accordingly of less relevance in explaining
the observed data.

Following \citet{ghir07b}, we also attempted ``constrained'' fits of
the BBCPL model in an attempt to force the black-body component to provide 
the peak of the spectral emission.  We fit a BBCPL model with the
black-body peak energy ($3.92~kT$) constrained to lie within the 90\% 
confidence limit of the best-fit Band \Ep\ value.\footnote{Our approach is
less restrictive than that of \citet{ghir07b}, who fixed the black-body peak
at \Ep.}  As expected, the
$\chi^2$ value for the constrained BBCPL fits increases dramatically relative to
the BBCPL fits, and the best-fit $kT$ value typically does not have 
90\% confidence limits within the allowed band.  
The black-body contribution to the total flux in the RHESSI band 
is not significantly different than for the regular BBCPL model 
and remains small.

\subsection{Time Evolution} \label{sec-evolution}
One of the most notable features of the BATSE BBPL fits presented by
\citetalias{ryde08} is the remarkably universal behavior of the time evolution obtained in
the BBPL fits.  For their sample of single-pulse bursts, they found that
both the fit black-body temperature and the bolometric black-body flux exhibit
a regular power law rise and decay.  They found identical break times for
the flux and temperature evolution in a given burst.  Moreover, a
dimensionless proxy for
the black-body normalization, $\Rt \equiv (F_{BB}/\sigma T^4)^{1/2} = 1.01
\times 10^{-16} A^{1/2} E_{piv}^{-1}$, is
related to the size of the photospheric emission region and 
increases monotonically even for bursts with complex light
curves.

We consider the time evolution of the black-body components of the RHESSI
BBPL and BBCPL fits in an attempt to reproduce these behaviors.  
Since the \citetalias{ryde08} time-evolution results were obtained from 
bursts with simple
single-pulse time profiles, we used a single-pulse burst sample with
slightly lower signal-to-noise ratio than our high S/N sample of mostly
complex bursts.  Unfortunately, only 3 bursts met the selection criteria
described in Section \ref{sec-sample}.  These bursts, GRBs 
020715, 030329A, and 060805B, have 8, 12, and 5 subintervals, respectively.
Accordingly, caveats of small sample size and limited time resolution
apply to our analysis of their time evolution.

\begin{figure} 
\plotone{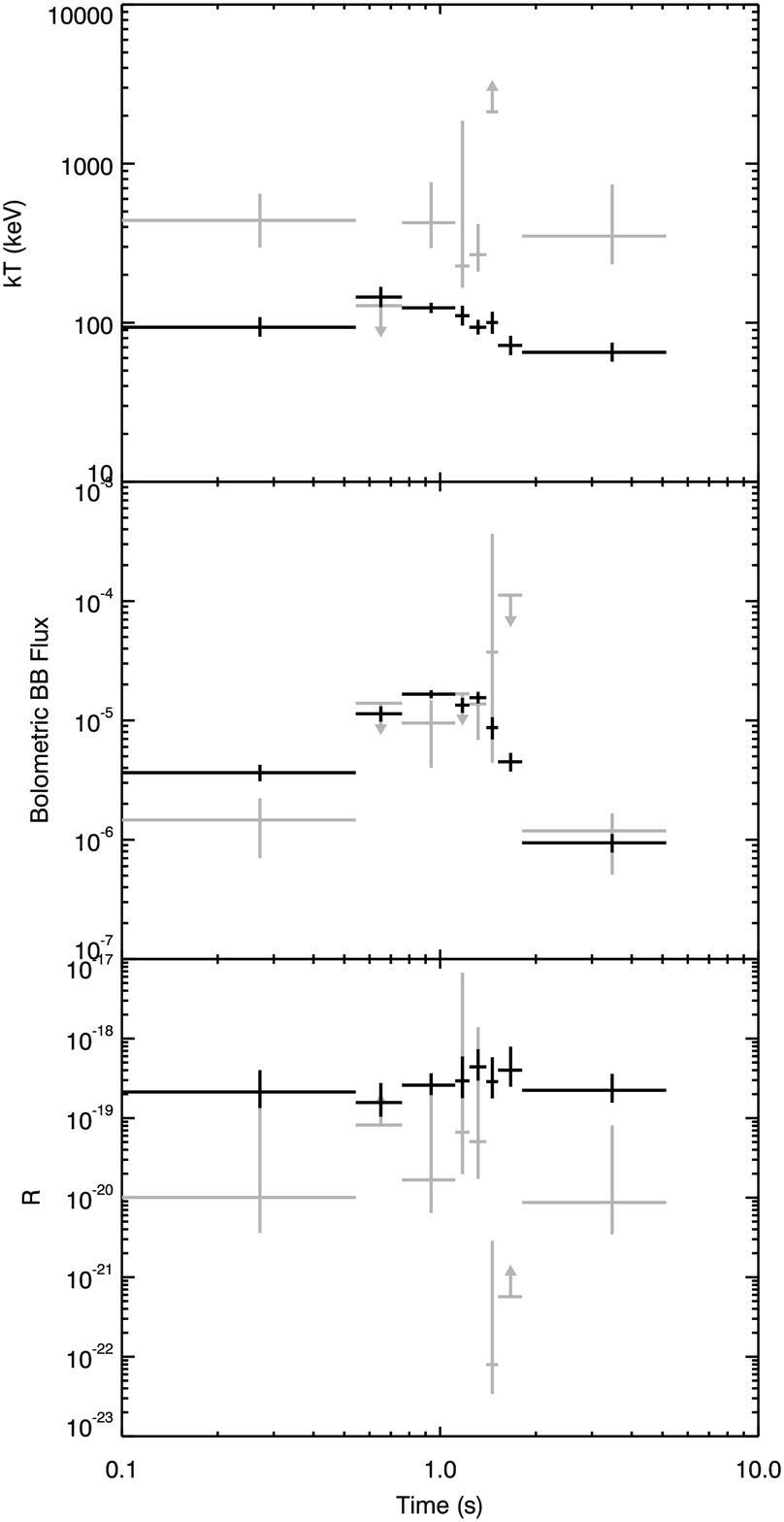}
\caption{
Time evolution of the temperature, black-body flux, and \R\ for the
single-pulse burst GRB 020715.  BBPL model fits are black, while BBCPL fits
are gray.
}
\label{fig-020715_log} 
\end{figure}

\begin{figure} 
\plotone{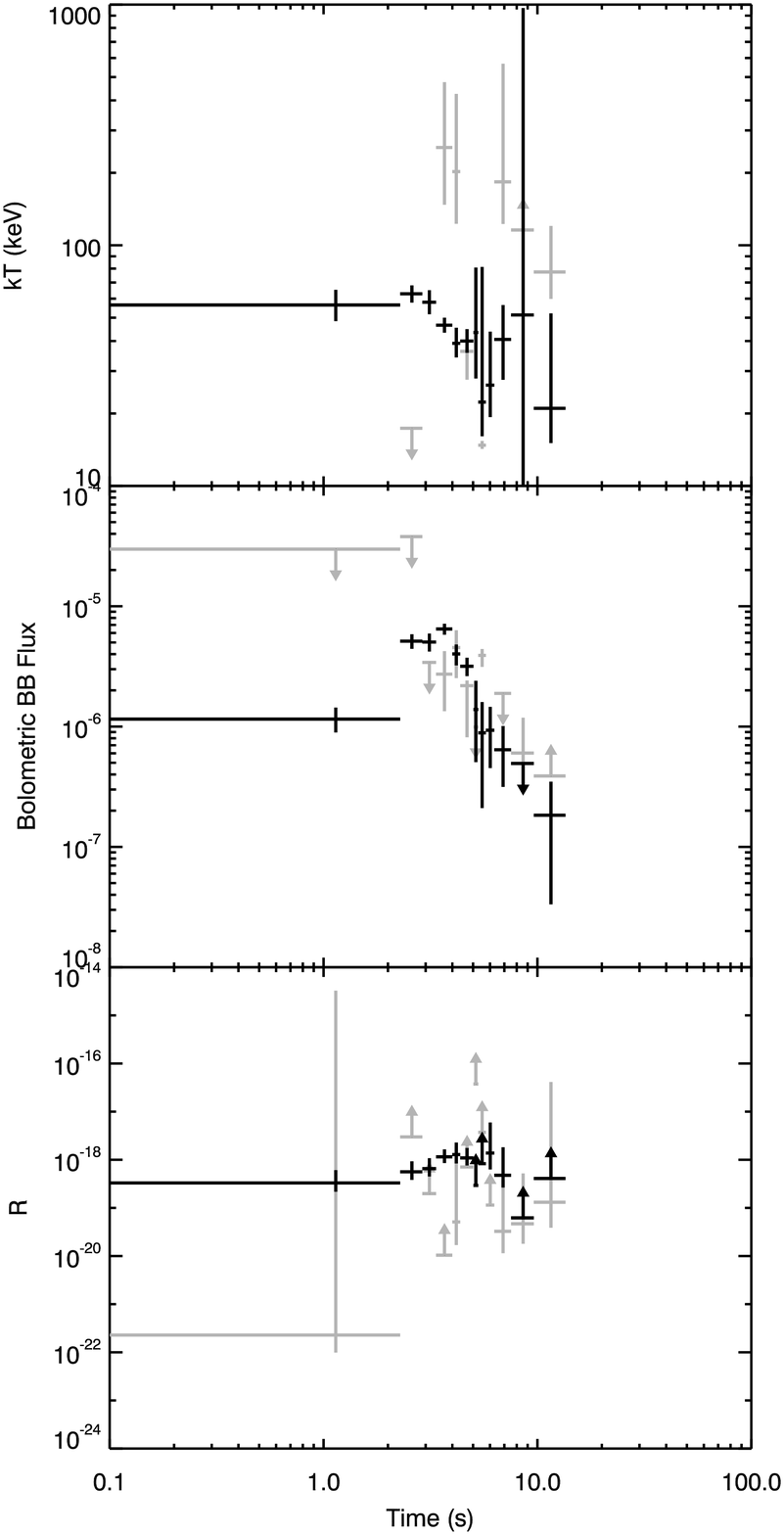}
\caption{
Data for the single-pulse burst GRB 030329.  Symbols as in Fig.
\ref{fig-020715_log}.
}
\label{fig-030329_log}
\end{figure}

\begin{figure} 
\plotone{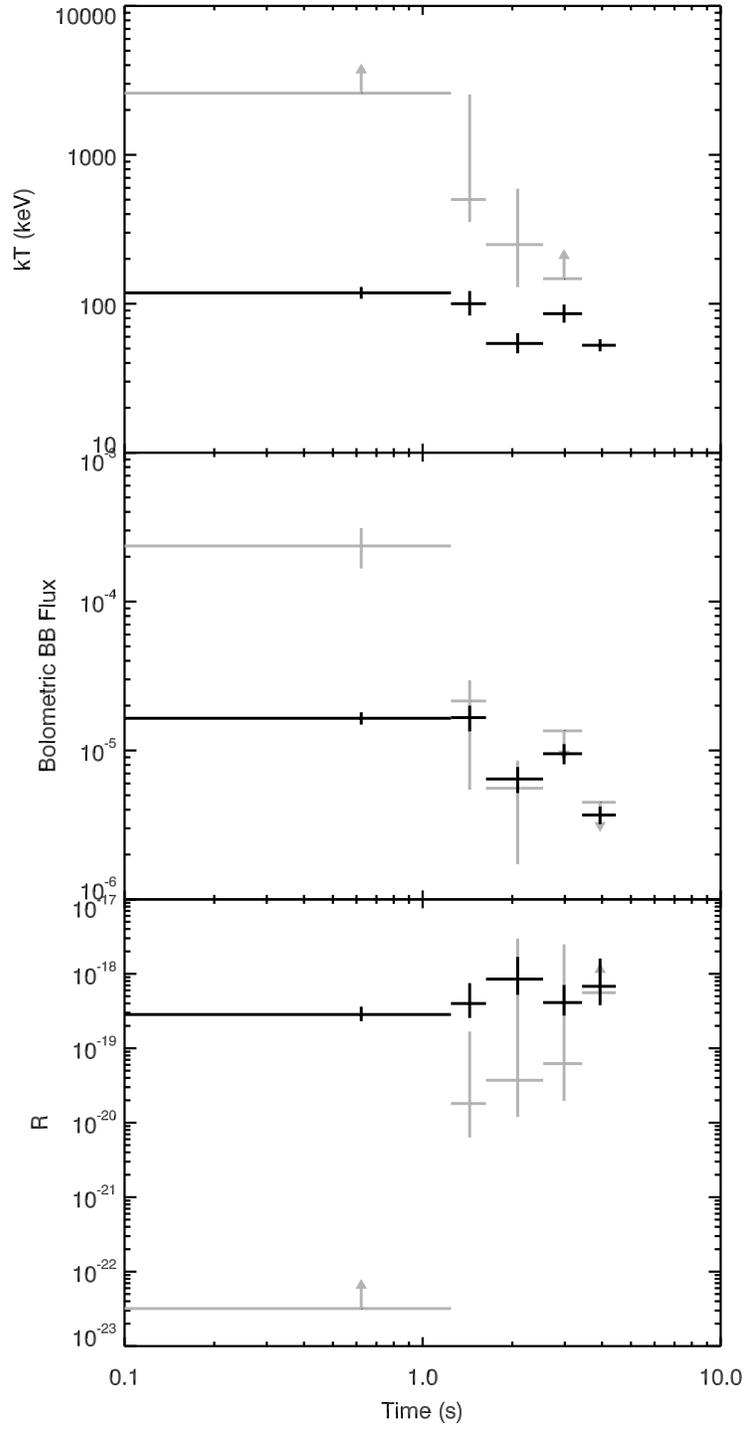}
\caption{
Data for the single-pulse burst GRB 060805.  Symbols as in Fig.
\ref{fig-020715_log}.
}
\label{fig-060805_log}
\end{figure}

We plot the time evolution of the temperature, bolometric black-body flux,
and \Rt\ in log-log space in Figures
\ref{fig-020715_log}, \ref{fig-030329_log}, and \ref{fig-060805_log}.
Within the moderate time resolution of these data, there is a clear
suggestion of the correlated power-law rise, break, and decay 
of the temperature and black-body flux for the BBPL model, as found by \citetalias{ryde08}.  

For these three bursts, the BBPL \Rt\ values are generally consistent with a
constant value.  While a minority of the \citetalias{ryde08} sample showed such behavior, far 
more frequently the \citetalias{ryde08} bursts had \Rt\ values which increased as
a power-law over an order of magnitude.  The authors interpreted this
increase in terms of the expansion of the photospheric emission region.
\citetalias{ryde08} also found frequent monotonic increases in \R\ for 
bursts with complex
lightcurves.  Our high S/N sample, consisting primarily of bursts with such
lightcurves, also all have \Rt\ consistent with a constant.  
The values of \R\
we obtain are consistent with the \citetalias{ryde08} sample, however.  Both the
single-peak and the high S/N samples range from $7 \times 10^{-20}$--$2
\times 10^{-18}$ and peak near $3 \times 10^{-19}$.

The time evolution of the BBCPL black body exhibits few clear
trends, as expected given the black body's convergence problems and minimal
contribution to the fit.
For the single-pulse sample, the bolometric black-body flux seems to track
that obtained in the BBPL model (Figures
\ref{fig-020715_log}--\ref{fig-060805_log}).  There is less
correlation between the temperatures of the two models, though, and the
time evolution of \Rt\ for both the single-pulse sample and the high S/N
sample shows little temporal regularity. 
Moreover, the BBCPL \Rt\ value can be more than an order of
magnitude different than that obtained in the BBPL fits, ranging from
$10^{-23}$--$10^{-17}$.  Such changes
would imply orders of magnitude change in the inferred size of photospheric
emission region, casting further doubt on the physical significance of the
black-body component of our BBCPL fits.

%%%%%%%%%%%%%%%%%%%%%%%%%%%%%%%%%%%%%%%%%%%%%%%%%%%%%%%%%%%%%%%%%%%%%%%%%%%%
\section{Conclusions}

We have tested quasi-thermal spectral models of GRB prompt emission over a
broader energy band (30~keV--17~MeV) than previously considered.  
Many of the successes of the BBPL model in the BATSE band (acceptable
$\chi^2_\nu$, physical spectral slopes for the nonthermal component,
universal time evolution) are not reproduced consistently in the RHESSI data
considered here.  We have attempted to construct more realistic
quasi-thermal models by combining a black body with a cutoff power-law
(BBCPL).  However, even using our intervals selected for high quality 
(S/N$ > 45$), the BBCPL model exhibits convergence difficulties.  More
crucially, though, the fit black-body component loses its relevance as the
origin of the spectral peak.  It is possible that bursts with simpler light
curves might exhibit better fit behavior or that data obtained over an even 
wider energy band could clarify the transitions between thermal and nonthermal
emission.  Data from {\it Fermi}-GBM and LAT will provide useful tests.
However, results from GRB 080916C, the first bright long burst observed by both
instruments, indicated that a Band function provided good
representation of the burst spectrum over the full 8~keV--200~GeV energy range
\citep{abdo09}.

The challenges of untangling the origins of the prompt GRB
emission from the prompt gamma-rays alone are long-standing.  
New observations of higher energy gamma-rays by {\it Fermi}-LAT, 
of simultaneous long-wavelength emission by transient monitoring campaigns, 
and of gamma-ray polarization by Compton telescopes 
will enrich our view of the prompt 
emission and provide new clues to its origin.

\section{Acknowledgements}
\acknowledgements 
ECB thanks Felix Ryde and Asaf Pe'er for useful discussions 
at the 2008 Huntsville GRB Symposium.  He is also grateful to Claudia
Wigger and Mark Bandstra for assistance in developing these analysis methods 
and for comments on a draft of this work.

{\it Facilities:} \facility{RHESSI}

\clearpage

\bibliography{grb}
\bibliographystyle{apj}

%\clearpage

\begin{deluxetable}{lccl}
\tablecolumns{4}
\tablewidth{0pt}
\tabletypesize{\footnotesize}
\rotate

\tablecaption{\label{tab-intervals}
Intervals used for spectral fitting.  The intervals are quoted relative to the start time T$_0$.
}

\tablehead{\colhead{GRB} & \colhead{Off-Axis Angle} & \colhead{T$_0$ (UT)} &
	\colhead{Interval Edges (sec)} 
}

\startdata
\cutinhead{High S/N Sample}
020715 & 72$^\circ$  & 19:21:08.666 &  0.000, 0.356, 0.646, 3.336 \\ 
021008A & 50$^\circ$  & 07:01:01.399 &  0.000, 1.265, 1.985, 2.541,
2.727, 2.934, 3.142, 3.278, \\
&&& 3.542, 3.889, 4.054, 4.314, 4.438, 4.995, 8.153, 8.617 \\ 
021206 & 18$^\circ$  & 22:49:12.951 &  0.000, 1.983, 2.163, 2.267,
2.405, 2.626, 3.012, 3.176, \\
&&& 3.377, 3.487, 3.559, 3.629, 3.714, 3.809, 3.871, 3.959, \\
&&& 4.024, 4.099, 4.361, 4.618, 4.699, 4.792, 4.945, 5.068, \\
&&& 5.168, 5.346, 5.552, 5.718, 6.112, 6.217, 6.603, 6.906 \\ 
030329A & 144$^\circ$  & 11:37:25.954 &  0.000, 2.266, 2.833, 3.520,
3.979, 4.351, 5.019, 6.226, \\
&&& 13.431, 14.335, 15.006, 15.687, 17.553 \\ 
030519B & 166$^\circ$   & 14:04:52.951 &  0.000, 5.769, 6.352, 6.726, 7.027, 7.214, 7.625, 11.065 \\ 
031027 & 102$^\circ$  & 17:07:37.744 &  0.000, 3.236, 4.774, 6.065, 9.469 \\ 
031111 & 156$^\circ$  & 16:45:20.776 &  0.000, 0.329, 0.971, 1.329 \\ 
040228 & 163$^\circ$  & 00:08:50.910 &  0.000, 4.751, 6.291, 8.458,
13.995, 15.009, 17.221, \\
&&& 17.563, 18.338, 22.981, 27.748 \\ 
040810 & 144$^\circ$  & 14:16:40.241 &  0.000, 17.055, 24.995, 45.637 \\ 

\cutinhead{Single-Pulse Sample}

020715 & 72$^\circ$  & 19:21:07.907 &  0.000, 0.543, 0.759, 1.115,
1.236, 1.397, 1.521, 1.810, \\
&&& 5.135 \\ 
030329A & 144$^\circ$  & 11:37:25.954 &  0.000, 2.280, 2.895, 3.356,
3.997, 4.355, 5.020, 5.287, \\
&&& 5.743, 6.287, 7.529, 9.616, 13.523 \\ 
060805B & 97$^\circ$  & 14:27:15.989 &  0.000, 1.245, 1.632, 2.537, 3.428, 4.451 \\ 
\enddata

\end{deluxetable}
 %\newpage
\begin{deluxetable}{llccccll}
\tablecolumns{8}
\tablewidth{0pt}
\rotate

\newcommand{\spm}[2]{\ensuremath{^{+#1}_{-#2}}}
\newcommand{\ppm}{\ensuremath{\pm}}
\tablecaption{\label{tab-fit_comps}
Comparison of time-integrated Band fit parameters for bursts fit in other works.  Results of the closest analagous fit are shown.
Only the limits of the RHESSI fit energy band are presented when different
ranges were used for the front and rear segments.  
Parameter uncertainties have been converted to 90\% C.L. where necessary.
}

\tablehead{\colhead{GRB} & \colhead{Instrument} & \colhead{Fit Band} & \colhead{$\alpha$} & \colhead{\Ep (keV)} & \colhead{$\beta$} & 
	\colhead{$\chi^2/$dof} & \colhead{Reference}
}

\startdata

020715 & RHESSI & 30~keV--17~MeV & -0.71\spm{0.11}{0.10} & 477\spm{59}{52} & -2.79\spm{0.24}{0.50} & 34.5/26 & (1) \\
 & RHESSI & 30~keV--15.7~MeV & -0.776\ppm0.072 & 531\ppm39 & -3.14\ppm0.41 & 106.3/113 & (2) \\
\tableline
021008A & RHESSI & 100~keV--17~MeV & -1.360\spm{0.035}{0.034} & 733\spm{16}{16} & -3.59\spm{0.11}{0.14} & 59.7/30 & (1) \\
 & RHESSI & 300~keV--15.7~MeV & -1.493\ppm0.092 & 677\ppm54 & -3.73\ppm0.30 & 79.5/84 & (2) \\
\tableline
021206 & RHESSI & 30~keV--17~MeV & -0.817\spm{0.031}{0.030} & 718\spm{17}{17} & -2.912\spm{0.053}{0.059} & 118/68 & (1) \\
 & RHESSI & 70~keV--4.5~MeV & -0.692\ppm0.033 & 711\ppm12 & -3.19\ppm0.07 & 176.5/174 & (2) \\
\tableline
030329A & RHESSI & 30~keV--17~MeV & -1.648\spm{0.060}{0.053} & 139\spm{7}{7} & -3.03\spm{0.24}{0.53} & 101.4/46 & (1) \\
 & RHESSI (peak 1) & 34~keV--10~MeV & -1.608\ppm0.063 & 157.2\ppm8.6 & -3.48\ppm0.87 & 84.3/90 & (2) \\
 & RHESSI (peak 2) & 34~keV--7~MeV & -1.78\ppm0.11 & 85\ppm18 & -3.04\ppm0.49 & 103.3/83 & (2) \\
 & HETE & 2~keV--400~keV & -1.32\ppm0.02 & 70.2\ppm2.3 & -2.44\ppm0.08 & \nodata & (3) \\
 & HETE & 2~keV--400~keV & -1.26\spm{0.01}{0.02} & 68\ppm2 & -2.28\spm{0.05}{0.06} & 213.7/139 & (4) \\
\tableline
030519B & RHESSI & 50~keV--17~MeV & -1.171\spm{0.049}{0.046} & 437\spm{23}{22} & -3.11\spm{0.23}{0.40} & 66/35 & (1) \\
 & RHESSI & 70~keV--15~MeV & -1.048\ppm0.069 & 417.2\ppm21 & -3.11\ppm0.30 & 86.3/75 & (2) \\
 & HETE & 2~keV--400~keV & -0.8\ppm0.1 & 138\spm{18}{15} & -1.7\ppm0.2 & 92.0/124 & (4) \\
\tableline
031027 & RHESSI & 30~keV--17~MeV & -0.759\spm{0.089}{0.080} & 319\spm{26}{24} & $<-2.79$ & 63.9/56 & (1) \\
 & RHESSI & 60~keV--6~MeV & -0.94\ppm0.13 & 338\ppm25 & \nodata & \nodata & (2) \\
\tableline
031111 & RHESSI & 30~keV--17~MeV & -1.071\spm{0.073}{0.061} & 841\spm{96}{89} & -2.83\spm{0.22}{0.38} & 50.2/56 & (1) \\
 & RHESSI & 38~keV--15.7~MeV & -1.102\ppm0.059 & 844\ppm97 & -2.364\ppm0.11 & 128.3/117 & (2) \\
 & HETE & 2~keV--400~keV & -0.82\spm{0.05}{0.5} & 404\spm{68}{51} & \nodata & \nodata & (5) \\
\tableline
040228 & RHESSI & 50~keV--17~MeV & -1.60\spm{0.037}{0.034} & 769\spm{120}{99} & -2.50\spm{0.18}{0.36} & 107.9/84 & (1) \\
\tableline
040810 & RHESSI & 30~keV--17~MeV & -1.45\spm{0.12}{0.10} & 321\spm{50}{45} & $<-2.52$ & 55.7/56 & (1) \\

\enddata

\tablecomments{(1) This work (2) \citet{wigg08} (3) \citet{vand04} (4) \citet{saka05} (5) \citet{pela08}} 
\end{deluxetable}
 %\newpage

\epsscale{0.55}

\epsscale{0.6}

\epsscale{1.0}

\epsscale{0.6}

\end{document}